# The chemisorption thermodynamics of $O_2$ and $H_2O$ on AFM $UO_2$ surfaces unraveled by DFT+U-D3 study

Yang Huang[a,†], Le Zhang[a,b,†], Hefei Ji[c], Zhipeng Zhang[a], Qili Zhang[a], Bo Sun[a,b,*], Haifeng Liu[a,b], Haifeng Song[a,b]

[a]*Institute of Applied Physics and Computational Mathematics, Beijing, 100094, China*

[b]*Laboratory of Computational Physics, Institute of Applied Physics and Computational Mathematics, Beijing 100088, China*

[c]*Institute of Materials, China Academy of Engineering Physics, Jiangyou, 621907, China*

---

[†] These authors contributed equally to this work.
[*] Corresponding author, sun_bo@iapcm.ac.cn

**Abstract**

Unraveling the adsorption mechanism and thermodynamics of $O_2$ and $H_2O$ on uranium dioxide surfaces is critical for the nuclear fuel storage and uranium corrosion. Based on the first-principles DFT+U-D3 calculations, we carefully test the effect of antiferromagnetic order arrangements on the thermodynamic stability of $UO_2$ surfaces and propose the 1k AFM surface computational model. The chemisorption states of $O_2$ and $H_2O$ on $UO_2$ (111) surface, suggested by previous experiments, are accurately calculated for the first time. The adsorption properties of $O_2$ and $H_2O$ on $UO_2$(111) and (110) surfaces are discussed in detail to reveal the different interaction mechanisms. Combined with *ab initio* atomistic thermodynamics method, we systematically calculate the chemisorption phase diagram and isotherm of $O_2$ and $H_2O$ on $UO_2$ surfaces. Due to the different intermolecular interactions, the monolayer and multilayer adsorption models are identified for $O_2$ and $H_2O$, respectively. This study has comprehensively revealed the different adsorption mechanisms of $O_2$ and $H_2O$ on $UO_2$ surfaces, bridging the electronic structure calculations to the interpretation of experimental results and providing a solid foundation for future theoretical studies of uranium corrosion mechanism in humid air.

## 1. Introduction

In the nuclear industry, it is a historical concern of special note to unveil the adsorption and reaction of $O_2$ and $H_2O$ on uranium dioxide ($UO_2$) surface, which is directly related to uranium corrosion failure and nuclear fuel storage [1–5]. Since 1950s, a large number of experimental studies on the surface corrosion chemistry of $UO_2$ have obtained rich and interesting corrosion behaviors and rules, such as the $H_2O$ accelerated dry-oxidation [6], the $O_2$ inhibited wet-oxidation [1], the $H_2$ releasing from $H_2O$ reaction [7], and the reaction sequence of $H_2O$ and $O_2$ mixture [8]. The essential explanation of macroscopic experimental phenomena has always attracted the exploration of microscopic mechanisms. Since 2009, the ground state of antiferromagnetic (AFM) Mott insulator $UO_2$ with strongly correlated 5f electrons can be accurately described by the DFT+U calculation [9], it has attracted extensive theoretical research attentions to the molecular adsorption and reaction mechanisms on $UO_2$ surfaces [10–16]. Although the different adsorption properties of $H_2$, $O_2$ and $H_2O$ have been found to explain the corresponding experiments [17–19], experimental and theoretical studies have not yet reached a complete and consistent picture of the adsorption thermodynamics on $UO_2$ surfaces.

From the perspective of practical application, the stable surface orientations and the ambient conditions should be considered to make a comprehensive comparison. According to the experimental result [20] and Tasker's theoretical calculations [21] on the surface stability of ionic crystal, the (111) surface orientation is the most stable for $UO_2$ with a fluorite-type structure, and the (110) surface second. The polar (001) surface is inherently unstable due to an alternating distribution of U and O ions. As shown by the characteristic Wulff shape of $UO_2$, the (111) surface is the dominant share, thus it is particularly so in surface corrosion chemistry [20]. Existing experiments have shown the different adsorption behaviors of $H_2$, $O_2$ and $H_2O$ on $UO_2$ surfaces, resulting in the complex corrosion behavior in the mix gas atmosphere [22,23]. Senanayake *et al*. [24] investigated $H_2O$ adsorption on stoichiometric and reduced $UO_2$ (111) surfaces, and obtained the corresponding desorption temperatures of the chemisorbed $H_2O$ (i.e. 400 K and 530 K). Danon *et al*. [17] identified different desorption peaks of $H_2O$ corresponding to the molecular and dissociative chemisorption states, which were also observed by Cohen *et al* [25]. Roberts *et al*. [19] pointed out that $O_2$ reacts with one $U^{4+}$ site to form a rapid chemisorption on the stoichiometric $UO_2$ surface, and the adsorption isotherm follows single-layer adsorption model. Ferguson *et al*. [26] found the adsorption heat of $O_2$ decreases significantly as its coverage increases. Compared with $O_2$ and $H_2O$, there are less experimental studies of $H_2$, focusing on the hydrogen generation during the U-$H_2O$ corrosion reaction [27,28]. Bloch *et al* [29]. pointed out that high temperature will cause the desorption of water and accelerating the adsorption of hydrogen. In order to elucidate the distinct adsorption properties of $H_2$, $O_2$ and $H_2O$, many theoretical calculations have discussed the underlying adsorption mechanism. Nevertheless, the effects of environmental temperature and gases pressure have not been thoroughly explored. Bo *et al*. [14,15] investigated the adsorption and dissociation of $H_2O$ molecules on $UO_2$ surfaces, and revealed that $H_2O$ molecules of high coverage tend to exhibit molecular adsorption or a combination of molecular and dissociative adsorption. Tegner *et al*. [11,12] pointed out that the surface orientation of $UO_2$ affects the adsorption energy of $H_2O$. Recently, Arts *et al*. [30] calculated the adsorption properties of $O_2$ and $H_2O$ on $UO_2$ surfaces, but found that $O_2$ undergoes physisorption on $UO_2$ (111) surface. Pegg *et al*. [31] examined the interactions between $H_2$ and $UO_2$ surfaces, and point out that $H_2$ is physically adsorbed on $UO_2$ (111) surface. One can see that current calculations have well described the experimentally observed $H_2O$ chemisorption, but have not captured the experimentally suggested chemisorbed state of $O_2$ on $UO_2$ (111) surface. At the moment, the exact mechanism on different electronic properties and thermodynamic pictures of $O_2$ and $H_2O$ chemisorption have not been obtained.

In terms of basic theoretic research, the AFM ground-state of the $UO_2$ is one of key factors to propose the reliable computational models [32,33]. The magnetic ordering of bulk $UO_2$ has been identified as 1k and 3k AFM by previous experiments [34,35]. Recently, the magnetic ordering of bulk $UO_2$ has been discussed based on the DFT+U calculations [36], and the 1k AFM ground-state is found to be slightly more stable than 3k AFM order. However, in order to avoid the complications for surface model design resulted from the AFM order, many theoretical studies [14,15] still use the ferromagnetic (FM) order. The impact of magnetic order arrangements on the accuracy of AFM $UO_2$ surface calculations has not been fully explored. For a large-scale computation model of AFM $UO_2$ surface, the 1k AFM order is undoubtedly one suitable choice, which requires the careful testing and verification.

In this work, we perform extensive first-principles calculations and *ab initio* atomistic thermodynamic study to investigate the adsorption behaviors of $H_2$, $O_2$ and $H_2O$ on AFM $UO_2$ (111) and $UO_2$ (110) surfaces, aiming at: (1) searching for the reliable computational model for AFM $UO_2$ surfaces; (2)

clarifying the different chemisorption states and mechanisms of $O_2$ and $H_2O$ molecules, as well as the physisorption of $H_2$; (3) unraveling the different chemisorption thermodynamics of $O_2$ and $H_2O$ from the perspective of practical application research, such as the long-term storage of nuclear fuel.

## 2. Computational method

### 2.1 First-principles calculation

The interactions between molecules ($O_2$, $H_2$ and $H_2O$) and $UO_2$ surfaces have been studied with the Vienna *ab initio* simulation package (VASP) [37]. The projector augmented wave (PAW) method is used to describe the electron interactions, and the electron exchange correlation energy is calculated within the generalized gradient approximation of the Perdew-Burke-Ernzerh of formalism [38]. A plane wave basis set with a cutoff energy of 500 eV is used. The Monkhorst−Pack k-point meshes of 6 ×6 × 6 and 6 × 6 × 1 are used for the calculations of $UO_2$ bulk and surfaces, respectively. The strong onsite Coulomb repulsion among the U-5f electrons is described with the DFT + U scheme formulated by Dudarev *et al* [39,40], and the $U_{eff}$ is set to 4 during the calculation [14]. DFT-D3 method is used to describe the van der Waals correlation between molecules and $UO_2$ surfaces [41,42]. The atom positions are relaxed until the residual force on each atom is less than 0.01 eV/Å. The calculated lattice constant of $UO_2$ bulk is 5.52 Å close to experimental value of 5.47 Å [43].

Since the electronic density of states (DOS) of bulk FM $UO_2$ is calculated to be unreasonably metallic (shown in Fig. S1), which is inconsistent with its typical Mott insulating nature. Therefore, the AFM ground-state of the $UO_2$ is considered in DFT+U calculations. In order to explore the effect of AFM order arrangements on $UO_2$ surface calculations, three crystal cells of the periodic 1k AFM $UO_2$ are constructed, with the longitudinal AFM arrangements along the [111], [110], and [001] orientations, respectively, as illustrated in Fig. S2. After structural optimization, the total magnetic moment of three cells all converges to zero, confirming the AFM nature. The total energies of the unit $UO_2$ from the three cells above are also close with subtle differences less than 0.2 eV, and the [001] oriented AFM $UO_2$ is slightly preferred, indicating that the magnetic order setting has little effect on the energy calculation of 1k AFM $UO_2$ bulk.

Furthermore, the (111), (110), and (001) surfaces of the AFM $UO_2$ bulk are constructed, with corresponding 1k AFM arrangement perpendicular to the surface. These surface structures also keep the AFM ground-state after structural optimization. The calculated surface energies of three $UO_2$ surfaces are presented in Table 1, indicating that the $UO_2$ (111) surface is the most stable, and the surface stability follows the order: (111) > (110) > (001), which is also consistent with experimental studies [10, 14, 44, 45].

**Table 1** The surface energy of AFM $UO_2$ (100), (110) and (111) surface. The literature results are by Arts [30] and Bo [14]

| Surface orientation | Surface energy (J/m$^2$) | Literature results (J/m$^2$) |
| --- | --- | --- |
| 111 | 0.60 | 0.72 [30], 0.73 [14] |
| 110 | 1.02 | 1.08 [30], 1.03 [14] |
| 100 | 1.47 | 1.52 [30], 1.47 [14] |

Recently, Arts *et al.*[30] have discussed the non-collinear 3k AFM order and the spin orbital coupling (SOC) effect in the DFT+U calculation of $UO_2$ surface, and pointed out that performing non-collinear calculations of surface diffusion or reaction remains practically impossible due to the computational cost. Actually, from a perspective of the applied research, the computational model including the SOC and 3k AFM order does not substantially improve the calculation accuracy of the $O_2$ and $H_2O$ adsorption, because the chemisorption state of $O_2$ has not been found. In fact, all current DFT+U calculations [30,36], including our work, consistently obtain the 1k AFM ground state of $UO_2$. Our work shows that the 1k AFM surface model, consisted of four $UO_2$ layers, is expected to be reliable to study the adsorption and reaction mechanisms on $UO_2$ surface. The AFM $UO_2$ (111) and (110) surface models for the following calculations of $H_2$, $O_2$ and $H_2O$ adsorption are shown in Fig. 1. The surface areas of $UO_2$ (111) and (110) are 52.74 Å$^2$ and 43.06 Å$^2$, respectively.

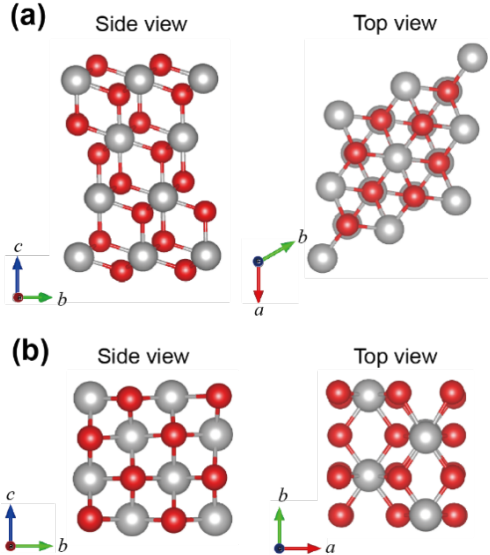

**Fig. 1** Side and top view of (a) $UO_2$ (111), and (b) $UO_2$ (110) surfaces. The red

and pewter balls represent O and U, respectively.

**2.2 *ab-initio* atomistic thermodynamics**

In order to discuss the adsorption thermodynamics on UO$_2$ surfaces under varying atmospheric pressures and temperatures, the *ab-initio* atomistic thermodynamic method [46] is utilized to calculate the surface energies of UO$_2$, which can be expressed as follows, [47,48]

$$\Gamma_{(i)} = \Gamma_{(i)}^0 + \theta_{M(i)}\left(E_a^{M(i)} + \Delta\mu_M\right) \tag{1}$$

where, $\Gamma_{(i)}$ and $\Gamma_{(i)}^0$ ($i$ = 111,110) are the surface energies with and without molecular adsorptions on the surface ($i$), respectively. $E_a^{M(i)}$ ($M$ = O$_2$, H$_2$O) is the average adsorption energy of molecule $M$ on the surface ($i$). $\theta_{M(i)}$ is the coverage of adsorbed molecule $M$ on the surface ($i$), written as the ratio between the number of adsorbed molecules $n_M$ and the surface area $A_{(i)}$. $\Delta\mu_M$ is the change in chemical potential of molecule $M$.

The surface energies without adsorptions could be calculated as follows,

$$\Gamma_{(i)}^0 = (E_{(i)}^N - NE_{UO_2}^{bulk})/(2A_{(i)}) \tag{2}$$

where, $E_{(i)}^N$ ($i$ =111,110) and $E_{UO_2}^{bulk}$ are the total energies of UO$_2$ slab and bulk, respectively.

The average adsorption energy ($E_a^{M(i)}$) in Eq. (1) could be calculated as follows,

$$E_a^{M(i)} = (E_{(i)+M} - E_i^N - nE_M)/n \tag{3}$$

where, $E_{(i)+M}$ is the total energy of surface ($i$) with $n$ adsorbed molecules, and $E_M$ is the total energy free molecule.

The change of chemical potential ($\Delta\mu_M$) in Eq. (1) is calculated by,

$$\Delta\mu_M = \mu_M^0(T,p_0) + k_BT\,ln(p_M/p_0) \tag{4}$$

where, $k_B$ and $T$ are Boltzmann constant and the temperature, respectively. $p_M$ and $p_0$ are the partial gaseous pressure of molecule $M$ and the standard state gaseous pressure. $\mu_M^0(T,p_0)$ is the standard chemical potential of molecular $M$ with temperature $T$ and gas pressure $p_0$, which could be expressed as follows,

$$\mu_0(T,p_0) = \Delta H_M(T,p_0) - T\Delta S_M(T,p_0) \tag{5}$$

where, $\Delta H_M(T,p_0)$ and $\Delta S_M(T,p_0)$ represent the changes of enthalpy and entropy of molecules $M$ at pressure $p_0$ when the temperature $T$ varies, respectively.

Based on Eqs. (1)-(5), the surface energies of UO$_2$ under various atmospheric conditions could be calculated to determine the adsorption phase diagram and then obtain the adsorption isotherm. As a result, the practical adsorption behaviors of different active molecules, and the corresponding adsorbability of different $UO_2$ surfaces could reasonably be predicted and estimated.

## 3. Results and Discussions

### 3.1 The individual adsorption properties of $O_2$, $H_2O$ and $H_2$ on $UO_2$ surfaces

#### 3.1.1 The adsorption on $UO_2$ (111) surface

First of all, the adsorption properties of a single $O_2$, $H_2O$ and $H_2$ molecule on the stable $UO_2$ (111) surface are systematically analyzed. Among plenty of possible adsorption configurations of $O_2$, $H_2$, and $H_2O$ molecules on the $UO_2$ (111) surface (Fig. S3), the most stable adsorption are searched out, and the corresponding configurations and adsorption energies are summarized in Fig. 2 and Table 2, respectively.

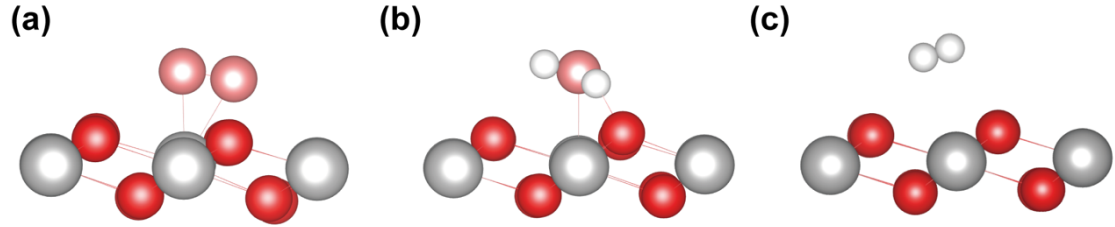

**Fig. 2** The stable adsorption configurations of (a) $O_2$, (b) $H_2O$ and (c) $H_2$ on $UO_2$ (111) surface. The pink and white represent the O and H atom in the molecules, respectively.

**Table 2** The adsorption energies of $O_2$, $H_2O$ and $H_2$ on $UO_2$ (111) and (110) surfaces (in eV/molecule).

| Surface orientation | $O_2$ | $H_2O$ | $H_2$ |
|---|---|---|---|
| 111 | -0.40 | -0.75 | -0.08 |
| 110 | -1.01 | -1.24 | -0.34 |

Fig. 2(a) shows that $O_2$ prefers to bond with one U atom on the (111) surface and its adsorption energy is calculated to be -0.4 eV, which is more negative or exothermic than the value -0.24 eV of the physisorption adsorption obtained by Arts *et al.* [30]. The more stable adsorption state of $O_2$ is searched out, and its adsorption configuration is close to the chemisorption state suggested by experiment [19]. Thus, the partial electronic density of states (PDOS) and the charge density differences are further compared and analyzed. The PDOS results of the free molecule state, the newly found stable state, and the reported physiosorbed state [30] are shown as (a) pre-adsorption, (b) chemisorption and (c) physisorption states in Fig.3, respectively. Evidently, when separated from UO$_2$ (111) surface by 7Å, the O$_2$ pre-adsorption preserves the orbital properties of free O$_2$ molecule, without the charge exchange with the surface. The PDOS of the physiosorbed O$_2$ in Fig. 3(c) is similar to the O$_2$ pre-adsorption, and Fig. 3(e) shows either no charge transfer with surface. In contrast, the charge density difference (Fig.3(d)) of our newly found stable state shows significant charge exchange with the surface. The corresponding PDOS in Fig.3(b) does verify that it is the chemisorption state of O$_2$ on UO$_2$ (111) surface. Specifically, after adsorption, the spin-up O-2p electrons occupied the $\pi_{2p}^*$ orbital of free O$_2$ which is obviously hybridized with the U-5f orbitals of UO$_2$ within the energy range of -2 to 0 eV. Meanwhile, the unoccupied anti-bonding $\pi_{2p}^*$ orbital of free O$_2$ undergoes splitting, and becomes the highest occupied molecular orbital (HOMO) of the adsorbed O$_2$.

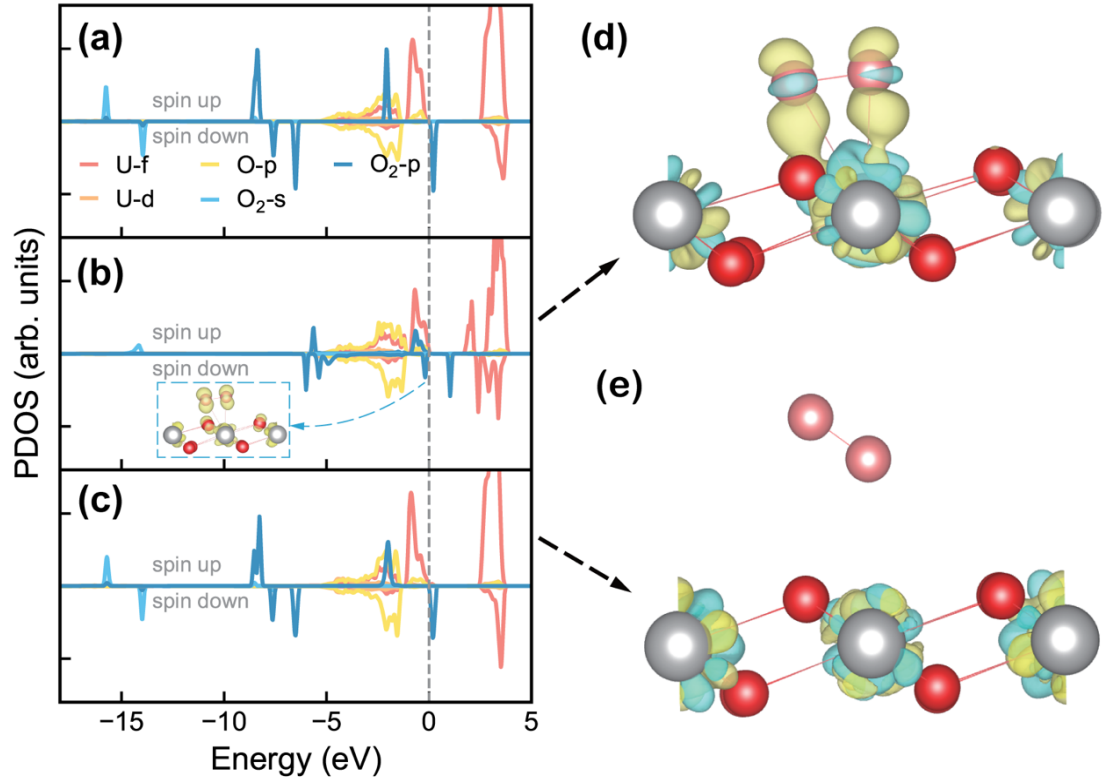

**Fig. 3** The PDOS results of (a) O$_2$ pre-adsorption, (b) O$_2$ chemisorption with insert showing partial charge density, and (c) O$_2$ physisorption [30] states, the Fermi level is set to zero. The charge density differences of (d) O$_2$ chemisorption and (e) physisorption states, where the yellow and cyan color represent gained and lost charge, respectively.

As illustrated in Fig. 2(b) and 2(c), the adsorptions of $H_2O$ and $H_2$ on $UO_2$ (111) surface are the molecular chemisorption and physisorption, respectively. The calculated adsorption energy of $H_2O$ is -0.75 eV, close to -0.73 eV reported by Arts *et al* [30]. The adsorption energy of $H_2$ is calculated only -0.08 eV, indicating the weak van der Waals interaction between $H_2$ and the surface. The different adsorption properties of $H_2O$ and $H_2$ are also well identified by the PDOS results in Fig. 4. Specifically, Fig. 4(a) shows that O-2p orbital of $H_2O$ tends to chemically bond with U-5f orbital, and the molecular orbital signature of free $H_2O$ disappears as a result of the charge exchange with the surface, which aligns with previous studies [11,14]. In contrast, $H_2$ molecules keeps the free molecule state without charge exchange with the surface after the physisorption.

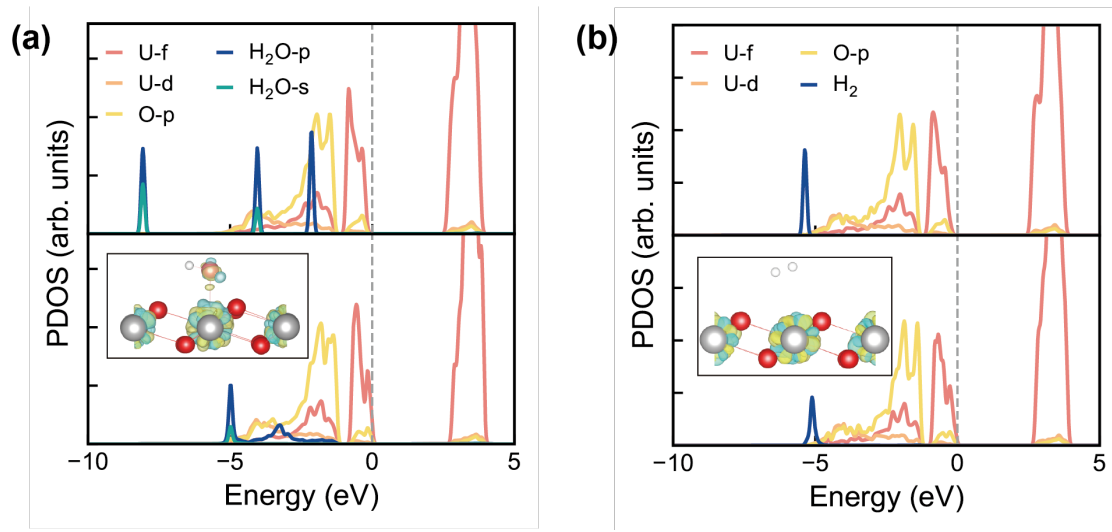

**Fig. 4** PDOS of (a) $H_2O$ and (b) $H_2$ adsorptions on $UO_2$ (111) surface: (top) pre-adsorption state and (bottom) post-adsorption states, with inserts showing the charge density difference, where the yellow and cyan colors represent gained and lost charge, respectively. The Fermi level is set to zero.

Given the above results, the adsorption properties of $O_2$, $H_2O$ and $H_2$ on the $UO_2$ (111) surface are can be briefly summarized as the moderate chemisorption of $O_2$, the strong chemisorption of $H_2O$, and the weak physisorption of $H_2$. As the most stable surface, $UO_2$ (111) will dominate the surface corrosion of $UO_2$ with a high probability. Thus, the accurate calculations of the chemisorption states of $O_2$ and $H_2O$, as well as the physisorption of $H_2$ have provided an important foundation towards the in-depth interpretation of the unclear experimental findings [1,6–8].

### 3.1.2 The adsorption on UO$_2$ (110) surface

In order to discuss the surface effect of orientation on the individual adsorptions of O$_2$, H$_2$, and H$_2$O on the UO$_2$ (110) surface are further studied. As shown in Table 2, all the adsorption energies of H$_2$, O$_2$ and H$_2$O on UO$_2$ (110) surface are much lower than on UO$_2$ (111) surface, suggesting that the adsorbability of UO$_2$ (110) surface is much stronger than that of UO$_2$ (111).

Fig. 5 presents results of the PDOS, adsorption configuration and charge density difference of the O$_2$, H$_2$, and H$_2$O adsorbed UO$_2$ (110) surfaces. The adsorption configurations of H$_2$ and H$_2$O on the (110) surface are similar to those on the (111) surface, i.e. a single molecule attached to one U atom. Whereas the chemisorbed O$_2$ prefers to bond with the two nearest U atoms on (110) surface, which differs from that on the (111) surface. In Fig. 5(a), the unchanged PDOS of H$_2$ indicate almost the same weak interaction with UO$_2$ (110) surface as with (111) surface. The unchanged bond length (0.74 Å) of H$_2$ also confirms the physisorption of H$_2$ on the UO$_2$ (110) surface. In contrast, the significantly changed PDOS of O$_2$ and H$_2$O in Fig. 5(b) and (c) indicates the typical chemisorption state. Moreover, the charge density differences in Fig. 5(e) and (f) show that the charge exchange on UO$_2$ (110) surface becomes more pronounced than that on UO$_2$ (111) surface. The partial charge density of HOMO for O$_2$ also verified the occupation of anti-bonding $\pi_{2p}^*$ orbital, which is consistent with that on UO$_2$ (111) surface.

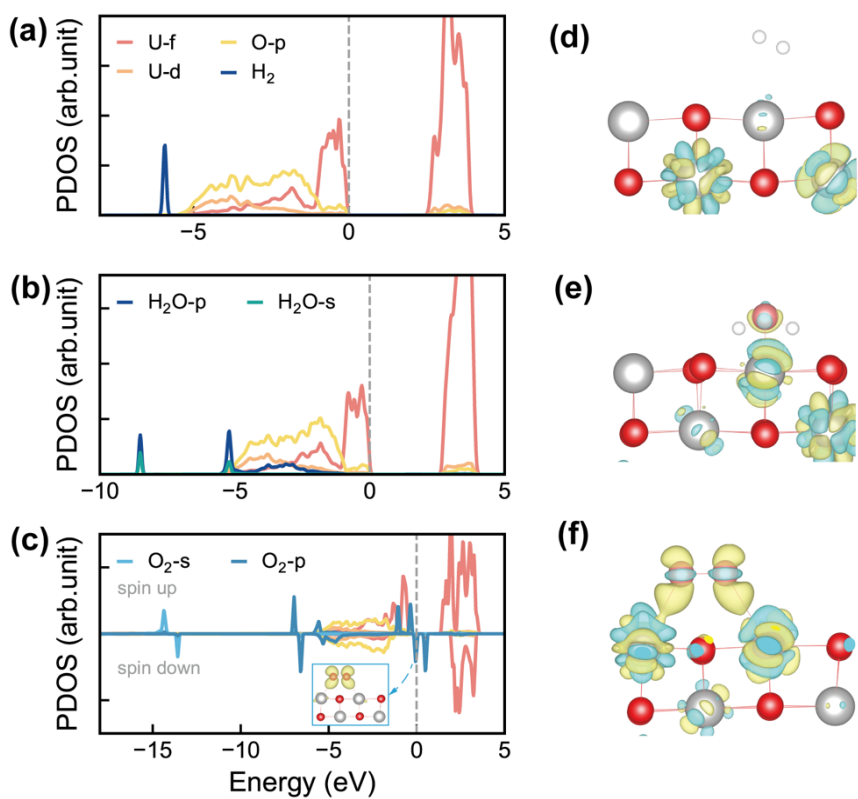

**Fig. 5** The PDOS of (a) $H_2$, (b) $H_2O$ and (c) $O_2$ adsorption with insert showing partial charge density, the Fermi level set to zero. The adsorption configuration and the charge density difference of (d) $H_2$, (e) $H_2O$ and (f) $O_2$ on the $UO_2$ (110) surface, where the yellow and cyan colors represent gained and lost charge, respectively.

As discussed above, the chemisorption of $O_2$ and $H_2O$ on the chemically active $UO_2$ (110) surface becomes more exothermic and stable than on relatively inert (111) surface. Unlike the oxidizing molecules, the adsorptions of reductive $H_2$ on stoichiometric $UO_2$ (111) and (110) surfaces only exist physisorption mainly via the weak van der Waals forces. All the theoretical studies to date have agreed on this point, which can help explain the macroscopic experiments, such as that $H_2$ has little or no effect on the U-$H_2O$ reaction [8]. Thus, in the subsequent sections, we mainly discuss the chemisorption mechanisms and behaviors of $O_2$ and $H_2O$ on $UO_2$ surfaces.

### 3.2 The coverage-related adsorption mechanisms of $O_2$ and $H_2O$ on the $UO_2$ surface

In order to obtain the basic mechanism for the experimentally observed different behaviors of $O_2$- and $H_2O$-adsorption on $UO_2$ surfaces under ambient pressure and temperature conditions, we further study the adsorption mechanism of multiple molecules, namely, discussing the variation of adsorption energies and configurations of $O_2$ and $H_2O$ with their coverage.

The stable adsorption configurations and adsorption energies of $O_2$ and $H_2O$ at different coverage on the $UO_2$ (111) surface are calculated and shown in Fig. 6(a). As the coverage increases, the preferred adsorption configuration of $O_2$ on the $UO_2$ (111) surface remains unchanged, but the adsorption configuration of $H_2O$ varies from the molecular adsorption to the mixture state of molecular and dissociated $H_2O$. This result aligns with the theoretical calculations reported by Tegner *et al* [11,12], and is consistent with the DSX experiments by Roberts *et al* [14].

By comparing the adsorption energies of $O_2$ and $H_2O$ at different coverage, it is evident that $H_2O$ consistently exhibits stronger adsorption on the $UO_2$ (111) surface than $O_2$, indicating that the adsorption of $H_2O$ is prior to $O_2$ in the humid air. As the coverage increased, the adsorption energies of $O_2$ rise, and the trend to become less exothermic is consistent with the experimental results by Ferguson *et al* [26]. However, the variation of the adsorption energy of $H_2O$ is not obvious with increasing coverage. The different trends of adsorption energies varied with coverage of $H_2O$ and $O_2$ may be due to the different interactions between the adsorbed molecules.

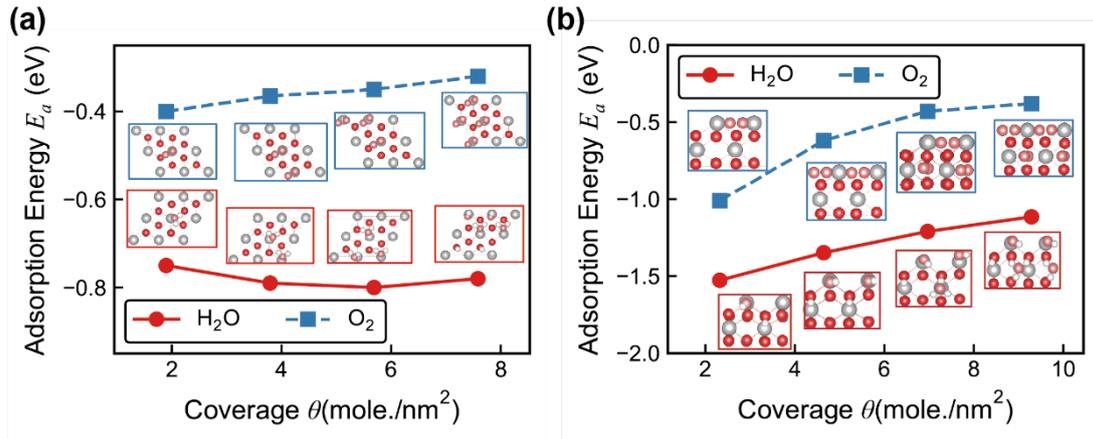

**Fig. 6** The adsorption energies ($E_a$) of $O_2$ and $H_2O$ varied with coverage ($\theta$) on (a) $UO_2$ (111) and (b) $UO_2$ (110) surfaces, with inserts showing corresponding stable adsorption configurations.

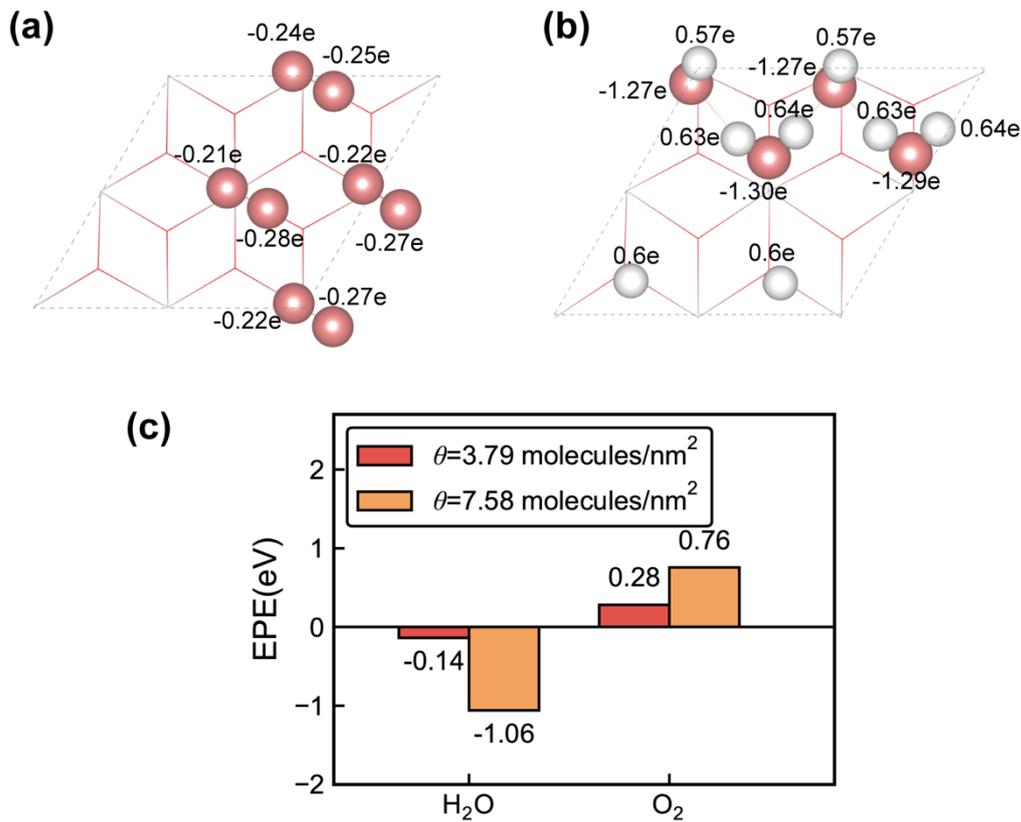

**Fig.7** The net charge of absorbed (a) $O_2$ and (b) $H_2O$ on the $UO_2$ (111) surface when coverage is 7.58 molecules/nm$^2$. (c) The average EPE of $O_2$ and $H_2O$ on the $UO_2$ (111) surface while coverage is 3.79 and 7.58 molecules/nm$^2$, respectively.

In order to elucidate the different intermolecular interactions of $O_2$ and $H_2O$, the net charge and Coulomb interactions of these molecules adsorbed on the $UO_2$ (111) surface are analyzed. Fig. 7(a) and 7(b) illustrate the net charge of $O_2$ and $H_2O$ on the $UO_2$ (111) surface. It is evident that each adsorbed $O_2$ molecule gains about 0.5e. In contrast, each H atom in molecular $H_2O$ and dissociative $H_2O$ loses about 0.6e, while O gains 1.3e. To quantitatively characterize the interactions between adsorbed molecules, the average electronic potential energy (EPE) of a single adsorbed molecule is calculated based on the net charge, employing methodologies detailed in our previous research [49]. As shown in Fig. 7(c), one can see there is repulsive interaction between adsorbed $O_2$ with negative charge, resulting in a positive EPE. In contrast, the calculated EPE for $H_2O$ is negative, indicating the attractive interaction between molecules. As the coverage increases, the absolute value of EPE increases for both $O_2$ and $H_2O$. The enhancement for $O_2$ is mainly due to the fact that stronger repulsive interaction between molecules at coverage increasing. The further decrease in the negative EPE for $H_2O$ could be due to the stronger attraction between molecules and hydrogen bonding between polar $H_2O$. Consequently, the different intermolecular interactions account for the diverse trends in adsorption energies of $O_2$ and $H_2O$.

In Fig. 6(b), we further discuss the effect of surface orientation by calculating the variation of adsorption energies of $H_2O$ and $O_2$ on the $UO_2$ (110) surface with surface coverage. Notably, $H_2O$ on the (110) surface tends to undergo dissociative adsorption at low coverage ($\theta =$ 2.32 and 4.64 molecules/nm$^2$), and becomes mixed adsorption states at higher coverage ($\theta =$ 6.97 and 9.29 molecules/nm$^2$). The adsorption energies and configurations of $H_2O$ at different coverage are similar with the existing research [14,30]. When the coverage increases, the adsorption energy of $O_2$ increase apparently, which is similar with $UO_2$ (111) surface. Unlike the decreasing trend of molecular $H_2O$ on (111) surface, the adsorption energies of $H_2O$ on (110) surface increase significantly with increasing coverage, which is mainly induced by the different stable adsorption configurations.

In conclusion, the distinct interactions in adsorbed $O_2$ and $H_2O$ result in significant differences in their adsorption behaviors. Due to the attractive interactions, $H_2O$ molecules tend to undergo multilayer adsorption. In contrast, $O_2$ molecules, experiencing repulsive forces, are inclined toward monolayer adsorption. Consequently, the thermodynamic adsorption behavior of $O_2$ and $H_2O$ under varying environmental conditions may differ markedly.

### 3.3 The adsorption thermodynamics of $O_2$ and $H_2O$ on $UO_2$ surface

In order to estimate the influence of environmental conditions on $O_2$ and $H_2O$ adsorptions on the $UO_2$ surfaces, the adsorption thermodynamics of $O_2$ and $H_2O$ are further analyzed based on *ab initio* atomistic thermodynamic calculations.

Fig. S4 (a) and (b) show the surface energies of the $UO_2$ (111) surface with $O_2$ and $H_2O$ adsorptions at various temperatures under one standard atmospheric pressure. $O_2$ undergoes chemisorption at temperatures below 180 K, with a surface coverage of 7.58 molecules/$nm^2$. As the temperature increases, the coverage of $O_2$ rapidly decreases, and $O_2$ molecules will completely desorb at 220 K. The previous experimentally observed $O_2$ chemisorption at the low temperature of -183°C [19] is eventually confirmed by our theoretical research.

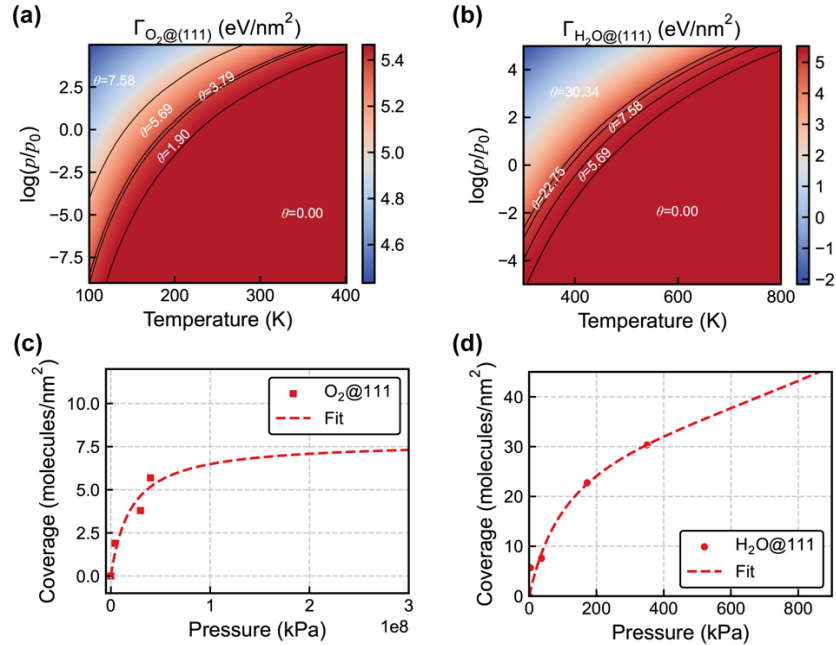

**Fig. 8** The surface energy of $UO_2$ (111) as a function of varying ($T$, $p$) with (a) $O_2$ and (b) $H_2O$ adsorption, respectively. The calculated isotherms for (c) $O_2$ and (d) $H_2O$ on $UO_2$ (111) surface at 400 K.

Under the standard atmospheric pressure, the effect of temperature on $H_2O$ adsorption behavior differs significantly from that of $O_2$. For $H_2O$ molecules, the surface coverage remains at a certain level at temperatures below 400 K. As the temperature increases, the coverage gradually decreases, reaching 5.69 molecules/$nm^2$ at 420 K. When temperature further increases, $H_2O$ eventually desorbs at 480 K. This desorption behavior is consistent with the two desorption peaks observed by the experiment [17], and the theoretical desorption temperature is also comparable to the experimental values (470 K and 570 K) when without considering the surface defects of the actual material.

Fig. 8 shows the adsorption phase diagrams of $O_2$ and $H_2O$ on the $UO_2$ (111) surface under varying pressure and temperature conditions. It can be seen that $O_2$ adsorption occurs under high pressure or low temperature. Under varying (*T, p*) conditions, the coverage of $H_2O$ is larger than that of $O_2$ at the same (*T, p*) before desorption, indicating that $H_2O$ molecules are more readily adsorbed on the $UO_2$ (111) surface than $O_2$. Based on the adsorption phase diagram, the adsorption isotherms of $O_2$ and $H_2O$ on the $UO_2$ (111) surface can be deduced, as shown in Fig. 8(c) and (d). Given the adsorption characteristics of $O_2$ and $H_2O$ discussed in Section 3.2, the Langmuir model ($\theta = a/(1+bp)$) and the BET model ($\theta = p/(ap^2 + bp + c)$) are used to fit the adsorption isotherms of $O_2$ and $H_2O$ molecules on the $UO_2$ (111) surface, respectively.

It can be seen in Fig. 8(c) that as pressure gradually increases, the coverage of $O_2$ molecules first increases rapidly and then stabilizes, consistent with the isotherm behavior of monolayer adsorption. The isotherm shape for $H_2O$ is significantly different from that of $O_2$. The coverage of $H_2O$ molecules continued to rise slowly and steadily with increasing pressure, consistent with the characteristics of multilayer adsorption. The goodness of fitting indicated that the equations achieved good fitting results, confirming that $O_2$ and $H_2O$ follow monolayer and multilayer adsorption behaviors, respectively.

In addition, we also calculate the surface energy of the (110) surface under atmospheric pressure as a function of temperature, as shown in Fig. 9. The results indicate that under atmospheric conditions, the adsorption characteristics of $O_2$ and $H_2O$ on the (110) surface differ from those on the (111) surface; they begin to adsorb at temperatures below 500 K and 600 K, respectively, which are significantly higher than those for the (111) surface. This suggests that the (110) surface exhibits higher adsorbability compared to the (111) surface.

Fig. 9 presents the adsorption phase diagrams of $H_2O$ and $O_2$ on the $UO_2$ (110) surface, and the adsorption isotherms of $H_2O$ and $O_2$ fitted by adsorption equations. Based on the comparison with Fig. 8, one can see that the gaseous pressure range for $O_2$ and $H_2O$ on the (110) surface is broader. This can be attributed to the higher adsorbability of the $UO_2$ (110) surface. The fitness of adsorption isotherms indicates that the adsorption of $O_2$ and $H_2O$ on the $UO_2$ (110) surface still following the monolayer and multilayer adsorption models, respectively.

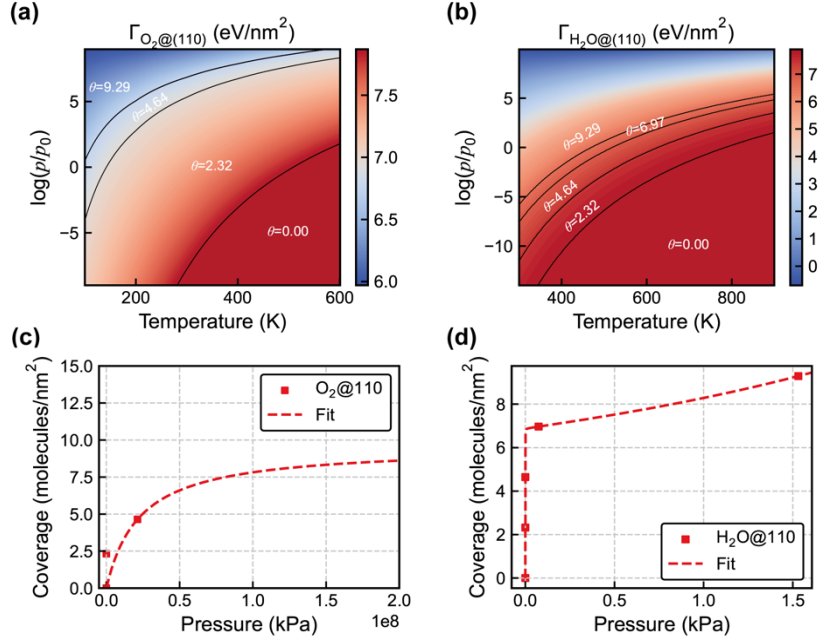

**Fig. 9** The surface energy of $UO_2$ (110) as a function of varying ($T$, $p$) with (a) $O_2$ and (b) $H_2O$ adsorption at different coverage, respectively. The calculated isotherms for (c) $O_2$ and (d) $H_2O$ on $UO_2$ (110) surface at 400 K.

Above all, the chemisorption of the polar $H_2O$ molecules is more readily and stable than $O_2$ on the stoichiometric $UO_2$ surfaces under the same pressure and temperature conditions. This is also the consensus of current theoretical research without considering the co-adsorption and the reaction of $H_2O$ and $O_2$. According to the experimental observation [22] that uranium corrosion in humid air firstly consumes $O_2$ and then $H_2O$, the corrosion reaction in mixed atmospheres does involve the complex interactions and has been calling for the further comprehensive investigations. Our current thermodynamic study of $H_2O$ and $O_2$ adsorption provides the solid foundation for the future studies of uranium corrosion in humid air.

## 4 Conclusion

Within the DFT+U-D3 scheme, extensive first-principles calculations combined with *ab initio* atomistic thermodynamics method are carried out to investigate the adsorption mechanisms and thermodynamic behaviors of $O_2$, $H_2$, and $H_2O$ on the AFM $UO_2$ (111) and (110) surfaces. The magnetic ordering is found to minimally affect the surface stability of $UO_2$, the reliable computational model of 1k AFM $UO_2$ film is thus proposed. the experimentally suggested chemisorption states of $O_2$ on $UO_2$ (111) surface are accurately captured, as well as $H_2O$. Whereas, $H_2$ cannot form a chemical bond and exchange charge with the $UO_2$ surface, showing the weak physical adsorption. In this work, the exact mechanism on different electronic properties and thermodynamic pictures of $O_2$ and $H_2O$ chemisorption have been obtained. With increasing coverage, the adsorption energies of $O_2$ significantly rise mainly due to the repulsive forces between absorbed $O_2$ molecules, whereas the adsorption energies of $H_2O$ are almost unchanged because of the attractive forces and existing hydrogen bonding interactions. The adsorption thermodynamics behavior indicates that $H_2O$ is more readily adsorbed on the $UO_2$ surface than $O_2$. Moreover, the $UO_2$ (110) surface exhibits higher adsorption activity than the $UO_2$ (111) surface. This study has enhanced the understanding of $UO_2$ surface interactions with complex environmental gases, providing new insights to bridge the electronic structure calculations to the macroscopic experiments.

**Acknowledgements**
This work was supported by National Nature Science Foundation of China, Nos. 12374056 and 11904027.